\documentclass[12pt]{article}
\usepackage{theorem}
\newcommand{\be}{\begin{equation}}
\newcommand{\ee}{\end{equation}}
\newcommand{\ie}{{\it i.e.}}
\newcommand{\g}{{\cal G}}
\newcommand{\fl}{}
\theorembodyfont{\rmfamily}
\newtheorem{definition}{Definition}[section]
\newtheorem{lemma}{Lemma}[section]
\newtheorem{example}{Example}[section]
\newtheorem{corollary}{Corollary}[section]

\begin{document}

\begin{flushright}
hep-th/9707032
\\
To appear in J. Phys. A
\end{flushright}

\vspace*{0.5cm}

\begin{center}

{\large\bf Generalized Jacobi structures}
\\[0.4cm]
{J C P\'erez Bueno
\footnote{On leave of absence from Departamento de F\'{\i}sica Te\'orica and 
IFIC (Centro Mixto Univ. de Valencia-CSIC) E--46100 Burjassot (Valencia), 
Spain. \\
E-mail: pbueno@lie.ific.uv.es}}
\\[0.3cm]
{\it Department of Applied Mathematics and Theoretical Physics, 
\\
Silver St., Cambridge, CB3 9EW, UK}

\end{center}

\begin{abstract}
Jacobi brackets (a generalization of standard Poisson brackets in which 
Leibniz's rule is replaced by a weaker condition) 
are extended to brackets involving an arbitrary (even) number of functions. 
This new structure includes, as a particular case, 
the recently introduced generalized Poisson structures.
The linear case on simple group manifolds is also studied and non-trivial 
examples (different from those coming from generalized Poisson structures) 
of this new construction are found by using the cohomology ring 
of the given group.
\end{abstract}

\section{Introduction}

Poisson structures (and Hamiltonian systems) can be introduced in geometrical 
terms by means of an appropriate bivector field $\Lambda$ verifying certain 
compatibility conditions that can be formulated by imposing the 
vanishing of the Schouten-Nijenhuis bracket (SNB) \cite{Sc,Ni} of $\Lambda$ 
with itself, $[\Lambda,\Lambda]=0$ \cite{Lich}.
This construction neither makes reference to symplectic structures 
nor requires a manifold of even dimension and provides a very convenient 
approach to generalize standard Poisson brackets.
Following this path, a generalization of standard Poisson structures has 
been introduced \cite{APPB} based on even multivector fields 
$\Lambda\in\wedge^{(2p)}$ having zero SNB with themselves
$[\Lambda,\Lambda]=0$. 
In the linear case, this new {\it generalized Poisson structure} (GPS) admits 
an infinity of examples related to the higher-order Lie algebras 
\cite{HIGHER}, a fact which generalizes the well known isomorphism between 
linear Poisson structures constructed out of the structure constants and 
(ordinary) Lie algebras.
The GPS are different from those proposed by Nambu long ago \cite{Na} where a 
(Nambu--)Poisson bracket 
involving three functions was introduced. 
Later Takhtajan \cite{Ta} extended the Nambu construction to a Nambu-Poisson 
bracket with an arbitrary number of functions (see also \cite{AG,ChaTak,Hiet}).

In this paper we construct a higher order generalization of the Jacobi 
structures \cite{Lichb,GL}, themselves a generalization of the standard Poisson 
structures, called local Lie algebras by Kirillov \cite{Kir}.
The generalization of the Poisson structures provided by the Jacobi ones is 
the result of substituting the Leibniz rule (derivation property) of the 
Poisson bracket by the weaker condition
\be
\hbox{support}\{f,g\}
\mathop{\subset}\limits_{\raise 0.3em\hbox{--}}
\hbox{support}\,f\cap\hbox{support}\,g\quad.
\label{conj}
\ee
Then, it is possible to show \cite{Kir} that the new bracket ({\it Jacobi 
bracket}) is a local type operator which has to be given by linear differential 
operators.
This implies that Jacobi structures,
in contrast with standard Poisson structures which may be determined uniquely 
by a bivector field $\Lambda$, 
are characterized by the differential 
operators defining the Jacobi bracket, namely a bivector and a vector 
fields $\Lambda$ and $E$.
If we want now the new bracket to satisfy the (standard) Jacobi identity (see 
(\ref{jacobi}) below), $\Lambda$ and $E$ must verify some compatibility 
conditions that can be expressed in terms of the Schouten-Nijenhuis bracket 
\cite{Lichb,GL}.
It is clear that {\it all} Poisson structures are also Jacobi structures 
because the Leibniz rule implies condition (\ref{conj}); this is the case when 
the vector field $E$ is set equal to 0. 

The aim of this paper is to show that, using the same geometrical approach 
by means of which (standard) Poisson structures can be extended to higher 
order GPS, Jacobi structures can also be 
extended to higher order {\it generalized Jacobi structures} (GJS).
In these, the generalized Jacobi brackets involve an arbitrary even 
number of functions.
They satisfy the same generalized Jacobi identity 
(GJI) 
introduced in \cite{APPB} (see (\ref{genjacident})) by virtue of which
both linear differential operators (a $2p$-vector and a $(2p-1)$--vector 
field) defining the generalized Jacobi bracket are constrained by some 
conditions expressed by means of the Schouten-Nijenhuis bracket.  
When the $(2p-1)$--vector field is set equal to zero we recover a standard 
Poisson structure (for $p=1$) or a GPS ($p$ arbitrary).
As a result, all GPS are also generalized Jacobi structures.
Although I have not been able to find a direct application of the GJS (which, 
as far as I know, is not easy even for the standard Jacobi structures),
I have been able to provide an infinite number of examples of these structures 
in the linear case, which extends greatly their mathematical interest.

The paper is organized as follows.
In Sec. \ref{secII} the definition of Jacobi bracket and Jacobi manifold is 
recalled \cite{Lichb,GL,Kir}.
In Sec. \ref{secIII} the GJS are introduced and some examples given.
Some conclusions close the paper.

\section{Jacobi manifolds}
\label{secII}

Let ${\cal F}(M)$  be the associative algebra of functions on the manifold $M$.

\begin{definition}[Jacobi bracket]
\label{def2.1}
A {\it Jacobi bracket} is a bilinear operation 
$\{\ ,\ \}:{\cal F}(M)\otimes{\cal F}(M)\to{\cal F}(M)$ which satisfies 
(\ref{conj}) and the following conditions $\forall f,g,h\in{\cal F}(M)$
\\[0.3cm]
a) skew-symmetry
\be
\{f,g\}=-\{g,f\}
\label{antisym}
\ee
b) Jacobi identity
\be
\{f, \{g, h\}\} + \{g, \{h,f\}\} +
\{h, \{f,g\}\}= 0\quad.
\label{jacobi}
\ee
\end{definition}

Conditions a) and b) endow ${\cal F}(M)$ with a structure of Lie algebra.
A manifold $M$ with a Jacobi bracket is called a Jacobi manifold.
If we substitute (\ref{conj}) for the stronger condition
\be
\{f,gh\}=g\{f,h\}+\{f,g\}h
\label{leibniz}
\ee
(Leibniz rule), we obtain a Poisson bracket (and then $M$ is called a Poisson 
manifold).

The more general form of a Jacobi bracket on the manifold $M$ is given 
\cite{Kir} by 
\be
\{f,g\}=\Lambda(df,dg)+fE(dg)-gE(df)\quad.
\label{jacobidef}
\ee
where $\Lambda$ and $E$  are, respectively, a two-vector and a vector field 
locally written as
\be
\Lambda={1\over 2}\Lambda^{ij}\partial_i\wedge\partial_j
\quad,\quad
E=\xi^i\partial_i
\quad.
\label{corrdinates}
\ee
Condition a) is automatically satisfied if $\{\ ,\ \}$ is defined by 
(\ref{jacobidef}).
Condition b) is taken into account by requiring 
\be 
[\Lambda,\Lambda]=2E\wedge\Lambda\quad,\quad [E,\Lambda]=0
\label{conditionb}
\ee
where $[\ ,\ ]$ stands for the Schouten-Nijenhuis bracket \cite{Sc,Ni}.
In fact (see \cite{Lichb})
\be
\fl
\epsilon^{ijk}_{123}\{f_i,\{f_j,f_k\}\}=
([\Lambda,\Lambda]-2E\wedge\Lambda)(df_1,df_2,df_3)-
\epsilon^{ijk}_{123}f_i[E,\Lambda](df_j,df_k)\quad,
\label{proof1}
\ee
so that, by requiring (\ref{conditionb}), the Jacobi identity is satisfied.
Thus \cite{Lichb}, a Jacobi structure on $M$ is defined by a 2-tensor 
$\Lambda$ and a vector $E$ satisfying the conditions (\ref{conditionb}).

It is clear that for $E=0$ we recover the equation
\be
[\Lambda,\Lambda]=0\quad,
\label{poissoncond}
\ee
which states that $\Lambda$ is a Poisson bivector and that $\{\ ,\ \}$ defines 
a Poisson structure \cite{Lich} on $M$.

In the same way that it is possible to characterize non-degenerate Poisson 
structures by covariant tensors satisfying $dF=0$, the Jacobi structures on a 
manifold of dimension $2n$ with non-degenerate bivector $\Lambda$ are 
characterized \cite{Kir,GL} by a two-form $F$ and a one-form $\eta$ 
which verify $dF=\eta\wedge F$ where $F$ and $\eta$ are 
given by their coordinates defined by
\be
\Lambda^{ik} F_{jk}=\delta^i_j\quad,\quad \eta_i=F_{jk} \xi^k\quad.
\ee

Examples of Jacobi structures (and Jacobi manifolds) are given by the 
locally conformal symplectic manifolds \cite{VaismanI} defined on an even 
dimensional manifold $M$ through a non-degenerate two-form $\Omega$ and a 
closed one-form $\omega$ (the Lee form \cite{Lee}) satisfying
\be
d\Omega=\omega\wedge\Omega\quad,
\ee
and the contact manifolds where we have a manifold $M$ with dim$\,M=2n+1$ and 
a one-form $\omega$ on $M$ (the contact form) which verifies
\be
\omega\wedge(d\omega)^n\ne 0\quad,\quad \forall x\in M\quad.
\ee

We want to recall here the linear case.

\begin{example}
\label{ex2.1}
Let $\Omega$ be the Poisson bivector associated with a Poisson-Lie structure 
(\ie, $\Omega={1\over 2} x_k C_{ij}^k\partial^i\wedge\partial^j$, 
where $C_{ij}^k$ are 
the structure constants of a Lie algebra $\g$);
then $[\Omega,\Omega]=0$. 
If we define the dilatation vector field 
$A=x_i\partial^i$, we may check that $[A,\Omega]=-\Omega$. 
So, defining $\Lambda\equiv\Omega +E\wedge A$ and imposing $[\Lambda,E]=0$ 
or, equivalently, $[E,\Omega]=-E\wedge [E,A]$ we obtain
\be
[\Lambda,\Lambda]=[E\wedge A,E\wedge A]+2[\Omega,E\wedge A]=
2E\wedge\Omega=2E\wedge \Lambda\quad;
\ee
hence, the pair ($\Lambda\equiv\Omega +E\wedge A\,,\,E$) defines a 
Jacobi structure if $[E,\Omega] = -E\wedge [E,A]$.

In particular, if $E$ is a constant vector, the condition above is equivalent 
to the one-cocycle condition for $E$, which reads
\be
\xi_\nu C^\nu_{ij}=0\quad.
\ee
For instance, if $\g$ is a simple (or semisimple) algebra the first cohomology 
group $H_1(\g)$ is zero (Whitehead's lemma), but we can take the algebra 
$\g\otimes u(1)$ for which $H_1(\g\otimes u(1))\ne 0$. 
Then, the bivector $\Lambda$ is given by 
\be
\Lambda= {1\over 2} x_k C_{ij}^k\partial^i\wedge\partial^j + 
x_i \partial^\varphi\wedge \partial^i\quad, 
\ee
where $\varphi$ denotes the coordinate corresponding to the $u(1)$ algebra 
generator (see \cite{dleon}). 
\end{example}
 
\section{Generalized Jacobi structures}
\label{secIII}

A natural higher order generalization of the standard Jacobi structures of 
Def. \ref{def2.1}
is given by a $2p$ and a $(2p-1)$-vector fields defining the 
linear mapping (cf. (\ref{jacobidef}))
\be
\fl
\{f_1,\dots,f_{2p}\}=\Lambda(df_1,\dots,df_{2p})-\sum_{j=1}^{2p}(-)^j f_j 
E(df_1,\dots,\widehat{df_j},\dots,df_{2p})\quad,\quad
\label{genjacdef}
\ee
which is antisymmetric in all its arguments $f_i$.
Then, to define generalized Jacobi structures we still have to impose a 
generalized Jacobi identity. This leads to

\begin{definition}[Generalized Jacobi structure]
\label{def3.1}
\quad A {\it generalized Jacobi structure} on the manifold $M$ is defined by  
a $2p$ and a $(2p-1)$-vector fields $(\Lambda,E)$ such that 
the mapping 
$\{\cdot,\dots,\cdot\}:{\cal F}(M)\times\mathop{\cdots}\limits^{2p}\times
{\cal F}(M)\to{\cal F}(M)$
given by (\ref{genjacdef}) 
satisfies the {\it generalized Jacobi identity} \cite{APPB} 
\be
\epsilon^{j_1\dots j_{4p-1}}_{1\dots 4p-1}
\{f_{j_1},\dots,f_{j_{2p-1}},\{f_{j_{2p}},\dots,f_{j_{4p-1}}\}=0\quad,\quad
\forall f_j\in {\cal F}(M)
\quad.
\label{genjacident}
\ee
The bracket (\ref{genjacdef}) will be called {\it generalized Jacobi bracket}.
\end{definition}

Now we need to characterize the generalized Jacobi structures in terms of 
the $2p$ and the $(2p-1)$-vector fields $(\Lambda,E)$.
This is achieved by the following

\begin{lemma}[Characterization of a GJS]
\label{lem3.1}
The linear mapping (\ref{genjacdef}) is a generalized Jacobi bracket (\ie, 
verifies (\ref{genjacident})) {\it iff} $\Lambda$ and $E$,
written in a local chart (cf. (\ref{corrdinates})) as
\be
\fl
\Lambda={1\over 2p!} 
\Lambda^{i_1\dots i_{2p}} \partial_{i_1}\wedge\dots\wedge\partial_{i_{2p}}
\quad,\quad
E={1\over (2p-1)!}
\xi^{i_1\dots i_{2p-1}} \partial_{i_1}\wedge\dots\wedge\partial_{i_{2p-1}}
\quad,
\ee
satisfy
\be
[\Lambda,\Lambda]=2(2p-1)E\wedge\Lambda\quad,\quad
[E,\Lambda]=0\quad.
\label{genconditionb}
\ee
\end{lemma}

\medskip
\noindent
{\it Proof:}\quad
The structure of the proof is equivalent to that for the standard 
$p=1$ case. 
In it we write the generalized Jacobi identity and factorize different kinds 
of terms. 
First we consider terms with first derivatives in $f$'s.
Those in (\ref{genjacident}) with the form 
$\partial f_1\dots\partial f_{4p-1}$ (all $f$'s derived once) 
are proportional to $((2p-1)(E\wedge\Lambda)-{1\over 2}[\Lambda,\Lambda])$.
Those with a non-derived $f$ are either proportional to $E\wedge E$ and hence 
directly zero ($E$ is of odd order) or proportional to $[E,\Lambda]$. 
Those with two non-derived $f$'s ($f_i$, $f_j$ say) are zero because they are 
symmetric under the permutation $f_i\leftrightarrow f_j$ while being 
antisymmetric in all the $f$'s the GJI.

The terms with second derivatives are proportional to
$$
\epsilon_{i_1\dots i_{4p-3}}(\Lambda^{i_1\dots i_{2p-1}\alpha}
\xi^{\beta i_{2p}\dots i_{4p-3}}+ 
\xi^{i_{1}\dots i_{2p-2}\beta}\Lambda^{\alpha i_{2p-1}\dots i_{4p-3}})\quad,
$$
or to
$$
\epsilon_{i_1\dots i_{n-1}\,j_1\dots j_{n-1}}
(\Lambda^{i_1\dots i_{n-1}\alpha}\Lambda^{j_1\dots j_{n-1}\beta}+
\Lambda^{i_1\dots i_{n-1}\alpha}\Lambda^{j_1\dots j_{n-1}\beta})\quad,
$$
which are zero being $E$ and $\Lambda$ of odd and even order respectively.
Thus, the unique conditions required to cancel all terms in the GJI 
are given by (\ref{genconditionb}), {\it q.e.d.}

\begin{corollary}
In the particular case $E=0$, 
(\ref{genjacdef}) reduces to 
$\{f_1,\dots,f_{2p}\}=\Lambda(df_1,\dots,df_{2p})$
and (\ref{genconditionb}) reduces to $[\Lambda,\Lambda]=0$, \ie, 
$\Lambda$ defines a GPS \cite{APPB}.
\end{corollary}

\begin{example}
\label{newexamp}
Let $M$ be a manifold with $\mbox{dim}\,M>2$; then if we take as $\Lambda$ a 
$(\mbox{dim}\,M)$-multivector field, for each $(\mbox{dim}\,M-1)$-vector $E$ we 
have a pair $(\Lambda,E)$ defining a GJS and $M$ becomes a generalized 
Jacobi manifold.
The conditions (\ref{genconditionb}) are satisfied because
$[\Lambda,\Lambda]$ and $E\wedge\Lambda$ are $(2\mbox{dim}\,M-1)$-vectors 
and $[\Lambda,E]$ is a $(2\mbox{dim}\,M-2)$-vector that are trivially zero on 
$M$. 
\end{example}

This is a very simple example that, in some sense, generalizes the fact that a 
two-vector on a two-dimensional manifold defines a (standard) Poisson 
structure.

\begin{example}
\label{ex3.1}
We can extend the linear example given in Sec. \ref{secII} to this case. 
To this aim let $\Omega$ be a $2p$-vector field defining a linear generalized 
Poisson structure (see \cite{APPB}), locally written as
\be
\Omega={1\over 2p!}\omega_{i_1\dots i_{2p}}^k x_k 
\partial^{i_1}\wedge\dots\wedge \partial^{i_{2p}}\quad,
\ee
and let $A$ be the dilatation operator as in Example \ref{ex2.1}.
Then, for every $(2p-1)$-vector field $E$ satisfying 
$[E,\Omega]=-E\wedge[E,A]$ (that is, $[E,\Omega+E\wedge A]=0$) we can 
define a generalized Jacobi structure given by the pair 
$(\Lambda\equiv\Omega+E\wedge A,E)$.
In particular, if $E={1\over (2p-1)!}\xi_{i_1\dots i_{2p-1}}
\partial^{i_1}\wedge\dots\wedge \partial^{i_{2p-1}}$ is a constant vector the 
condition on $E$ reduces to the expression
\be
\epsilon^{i_1\dots i_{2p-2} j_1\dots j_{2p}}_{k_1\dots k_{4p-2}}
\xi_{\nu i_1\dots i_{2p-2}}\omega_{j_1\dots j_{2p}}^\nu =0\quad,
\ee
or, equivalently,
\be
\partial_\Omega E=0
\ee
where $\partial_\Omega$ is the coboundary operator for the generalized Poisson 
cohomology introduced in \cite{APPB}. 
In contrast with the standard $p=1$ case, we do not need to `extend' the 
algebra to find $(2p-1)$-cocycles for the coboundary operator 
$\partial_\Omega$\footnote{This is an important difference with the standard 
$p=1$ case (Sec. \ref{secII}) in which we cannot define linear Jacobi 
structures on the dual of a simple Lie algebra.}.
In fact, as shown in \cite{APPB} (see also \cite{HIGHER,J.M.US}), 
all the higher order $\g$-cocycles for the 
ordinary Lie algebra cohomology are cocycles for the $\partial_\Omega$ 
cohomology. In other words, it is sufficient to find a simple Lie algebra 
with cocycles of orders $(2p-1)$ and $(2p+1)$ (or, in terms of the associated 
invariant polynomials, Casimirs of orders $p$ and $p+1$).
This is the case, for instance, for $su(3)$ where we find the generalized 
Jacobi structure given by the pair $(\Omega+E\wedge A, E)$ where
\be
\begin{array}{l}
\displaystyle
\Omega=
{1\over 4!} \epsilon^{j_2j_3j_4}_{i_2i_3i_4}d_{k_1k_2}^\sigma C^{k_1}_{i_1 j_2}
C^{k_2}_{j_3 j_4} x_\sigma \partial^{i_1}\wedge\partial^{i_2}
\wedge\partial^{i_3}\wedge\partial^{i_4}\quad, \\[0.45cm]
\displaystyle
E=
{1\over 3!} C_{i_1 i_2 i_3}\partial^{i_1}\wedge\partial^{i_2}
\wedge\partial^{i_3}\quad,
\end{array}
\ee
the coordinates $\xi_{ijk}=C_{ijk}$ of $E$ are the structure constants of 
$su(3)$ and the $d_{ijk}$ are the constants which appear in 
the anticommutators of the Gell-Mann matrices $\lambda$,
\be
\{\lambda_i,\lambda_j\}={4\over 3}\delta_{ij}1_3+2d_{ijk}\lambda_k
\quad.
\ee

The same construction extends to $su(l+1)\sim A_l$ ($l\ge 2$) for which we 
have $l$ primitive invariant polynomials of orders $2,3,\dots,l+1$ and hence 
$l$ cocycles of orders $3,5,\dots, 2l+1$.
Thus, for every cocycle (different from the first one of order 3 which defines 
the standard Poisson/Jacobi structure) we can give a non-trivial generalized 
Jacobi structure.
This explains why the standard case is singular and we have no linear 
Jacobi structures on the simple groups (defined by the 
tree-cocycle given by the structure constants which always exists). 
\end{example}

\section{Conclusions}

Despite the lack of a Leibniz rule that permits us to define a simple dynamics 
by $\dot f=\{H,f\}$ (where $\{\ ,\ \}$ stands for a Jacobi bracket) or, in the 
generalized case, $\dot f=\{H_1,\dots,H_{2p-1},f\}$ (see \cite{APPB} for a 
discussion on generalized Poisson dynamics) the Jacobi structures are not 
devoid of physical interest (and, of course, of mathematical one).

Generalized Poisson structures \cite{APPB} 
(see also \cite{Z2graded} for the $Z_2$-graded case) and their 
higher-order algebra counterparts \cite{HIGHER} provide a particular example 
of {\it strongly homotopy algebras} \cite{LAST,LAMA}
which are relevant in certain structures appearing in closed string theory and 
in connection with the Batalin-Vilkovisky formalism 
(see {\it e.g.}, \cite{ZIE,YKosmann}; for an account of the Batalin-Vilkovis\-ky 
formalism see \cite{MH,GPS}). 
It has been mentioned recently \cite{ILM} that there is a relation between 
Batalin-Vilkovisky algebras and Jacobi manifolds, although such a connection 
has not yet been made explicitly.
Clearly, the standard and the generalized Poisson structures \cite{APPB} are 
also special examples of the Jacobi structures considered here (it is 
sufficient to set $E=0$ and add the Leibniz rule) and, as such, they may share 
some properties, but more work is needed to analyze any physical applications 
of the GJS and, in particular, their possible quantization.
Note already that although (standard) Poisson brackets may be quantized by 
the bracket of associative operators that verifies the Leibniz rule 
$$
[A,BC]= ABC - BCA = [A,B]C + B[A,C]
$$
(as well as skewsymmetry and Jacobi identity) the standard Jacobi structure
does not satisfy this relation (unless it defines also a standard Poisson 
structure).
Moreover, in general, the skewsymmetrized product of an arbitrary (even) 
number of associative operators does not satisfy the Leibniz rule
(despite it verifies the generalized Jacobi identity \cite{HIGHER}).

From a purely mathematical (but nevertheless relevant) point of view, the 
mathematical contents (see Example \ref{ex3.1}) give to the new GJS a special 
interest, particularly in the linear case, where we have been able to provide 
examples associated with the cohomological properties of the Lie algebras.
This raises the question of whether other relations among the cocycles of a 
given Lie algebra may give rise to generalized Jacobi brackets.
This is matter for further work.

\subsection*{Acknowledgements}
This research has been partially supported by a research grant from the Spanish 
CICYT.
The author wishes to thank J. A. de Azc\'arraga for helpful discussions and 
comments on the manuscript.
The kind hospitality extended to him at DAMTP and an FPI grant from the 
Spanish Ministry of Education and Science and the CSIC are also gratefully 
acknowledged.


\end{document}